\documentclass[apsprd]{revtex4-1}
\usepackage{amsfonts}
\usepackage{amsmath}
\usepackage{amssymb}
\usepackage{graphicx}
\usepackage{color}

\def\be{\begin{equation}}
\def\te{\end{equation}}
\def\ee{\end{equation}}
\def\ba{\begin{eqnarray}}
\def\bea{\begin{eqnarray}}
\def\nn{\nonumber\\}
\def\tea{\end{eqnarray}}
\def\ea{\end{eqnarray}}
\def\eea{\end{eqnarray}}

\begin{document}

\title{Steady asymptotic equilibria in conformal relativistic fluids}

\author{Esteban Calzetta}
\email{calzetta@df.uba.ar}
\affiliation{Departamento de F\'isica, Facultad de Ciencias Exactas y Naturales, Universidad de Buenos Aires and IFIBA, 
CONICET, Cuidad Universitaria, Buenos Aires 1428, Argentina}


\begin{abstract}
When one considers a shock wave in the frame where the shock is at rest, on either side one has a steady flow which converges to equilibrium away from the shock. However, hydrodynamics is unable to describe this flow if the asymptotic velocity is higher than the characteristic speed of the theory. We obtain an exact solution for the decay rate to equilibrium for a conformal fluid in kinetic theory under the relaxation time approximation, and compare it to two hydrodynamic schemes, one accounting for the second moments of the distribution function and thus equivalent, in the small deviations from equilibrium limit, to an Israel-Stewart framework, and another accounting for both second and third moments. While still having a finite characteristic speed, the second model is a significant improvement on the first.
\end{abstract}
\maketitle


\section{Introduction}

Shock waves are one of the most interesting phenomena in relativistic hydrodynamics \cite{RZ13} and as such have spawned a significant literature \cite{Israel60,Taub73,Thompson86,CM88,MM90,MM93,KO93,Bouras09,Bouras10,MNR10,KKM10,Herbst19,RXZ20,LO11,ED18,HF21,Gabbana20}. In this work we want to focus on one aspect of the problem, which has been long considered critical for the development of viscous relativistic hydrodynamics \cite{I88}.

When one has a stationary shock wave, on both sides of the shock there is a steady flow which converges to equilibrium as we move away from the shock; the flow is supersonic on one side (which we shall define to be the left side) and subsonic on the other. This would seem to be a very simple configuration, nevertheless relativistic hydrodynamics is uncapable to describe it unless the asymptotic velocity, on the supersonic side, is below some threshold. The reason is that viscous relativistic hydrodynamics is build to transmit signals at a definite characteristic speed strictly less than that of light \cite{OH90,Muller99}; we show this explicitly in Appendix (\ref{signalprop}). We will comment further on why the characteristic velocity sets a limit for the existence of smooth solutions in the Results section \ref{results}. Similar problems arise already when studying shocks in non relativistic dilute gases \cite{WCU70,Sir63a,Sir63b,Strut08,J14}, see \cite{StrutBook}.

As a way out of this situation, we shall endorse the view that viscous relativistic hydrodynamics must be regarded not as a single theory but rather as a hierarchy of theories of increasing complexity. The more complex theories allow for faster signal propagation than the simpler ones, and so, although every single theory has a finite threshold, any shock wave in Nature may be described by a sophisticated enough theory \cite{MJ89,JP91}. 

In models whose fundamental description is kinetic theory, a particular way of building this theoretical hierarchy is by parameterizing the one particle distribution function in such as way that the parameterized distribution function reproduces the evolution of $N$ moments of the actual distribution function \cite{DMNR11,DMNR11b,DMNR12,DMNR12b}. In this class of models it may be proved that the fastest speed of propagation increases with $N$ and tends to the speed of light as $N\to\infty$ \cite{BR99}.

In this work we shall demonstrate a particular realization of this scenario. We shall consider a conformal fluid \cite{RR19} and we shall assume that its first principles description is given by kinetic theory under the relaxation time or Anderson-Witting approximation \cite{AW1,AW2,TI10,KW19,PC21}. We shall derive an exact expression for the decay constant of the solution away to equilibrium, and compare it with two hydrodynamic models of the divergence type theory (DTT) class \cite{L72,LMR86,GL90,GL91,PRC09,PRC10a,PRC10b,PRC13a}. The first is build to match the second moments of the distribution function, and the second is an improved version that also matches the third moments.

It should be noted that both these theories have some interest on their own. The first one has been used to analyze flows on Bjorken and Gubser backgrounds \cite{lucas19}
 and also the interaction between viscous fluids and gravitational waves in the Early Universe \cite{MGC17,MG21}. It has been extended to include thermal \cite{MGKC20} and turbulent \cite{EC21} fluctuations. It has also been extended to charged plasmas to study the amplification of magnetic fields in the Early Universe \cite{MGKC21}. By adding also the third moments, one obtains a theory which reproduces the propagators of the energy momentum tensor as derived from kinetic theory \cite{PC21}; it also recovers the dynamics of the spin $2$ degrees of freedom in the fluid as wave like, and not simply relaxational.

The paper is organized as follows. In next section we  fix our notation by considering shock waves in ideal \cite{RZ13}, Landau-Lifshitz \cite{LL6,KKM10}, and Israel-Stewart \cite{IS79a,IS79b,KKM10,MM85,DKKM08,U16} fluids. The Landau-Lifshitz framework does yield a finite decay rate for any asymptotic speed, but it seems to be an artifact beyond the limit  of weak shocks. When one regards hydrodynamics as rooted in kinetic theory, the Chapman-Enskog framework leads to the Landau-Lifshitz theory, and the Grad approximation to the Israel-Stewart one \cite{IS79a}. We shall show this connection in Appendix \ref{CHEap}. This means that the Israel-Stewart decay rate (with its limitations) will obtain in any theory that reduces to Grad's in the small deviations from equilibrium limit, such as anisotropic hydrodynamics \cite{ST14,KKL19,KL21} or our first DTT.

In section \ref{kt} we analyze the same problem within kinetic theory with an Anderson-Witting collision term. We show that there is a finite decay rate for any value of the asymptotic fluid velocity in the shock frame. That settles the issue that the problem of theory breakdown for strong shocks lies entirely within hydrodynamics. The dependence of the decay rate on the asymptotic velocity ressembles that derived from holography \cite{KKM10} but the divergence of the decay rate as the asymptotic velocity approaches light speed is stronger.

In section \ref{DTT} we consider the decay rate in our DTT. Since we already know the first DTT will revert to Israel-Stewart, the emphasis is on the second one, including third moments. This theory still has a highest propagation speed strictly less than light, and therefore also breaks down for a finite asymptotic velocity, but nevertheless it is a significant improvement on the Israel-Stewart result, both on the left and right sides of the shock.

We summarize our results and conclusions in the final section.

We have left some further details for the Appendixes. Appendix \ref{CHEap} shows the connection of the approaches in section \ref{common} to kinetic theory. The following two appendices have purely technical details. Appendix \ref{entropy} shows that consideration of the entropy current \cite{L08,JBP13,CJPR15} makes the dynamics of viscous relativistic fluids essentially unique. In Appendix \ref{signalprop} we compute the speed of signal propagation in both DTTs, thus allowing to check directly that it is the speed of propagation that defines the maximum asymptotic velocity the theory can handle \cite{OH90}, and finally in Appendix \ref{rightside} we shall discuss the straightforward modifications of our argument to compute the decay rates in the subsonic side of the shock.

\section{Common approaches to relativistic fluids}
\label{common}

\subsection{Shocks in ideal fluids}

An ideal fluid may be at equilibrium at each side ($L,R$) of the shock, with a discontinuity in temperature and velocity accross the shock. We assume the shock lies at the $z=0$ plane and is isotropic and translation invariant in this plane, and that all quantities depend only on the distance to the shock $z$. The discontinuity is restricted by EMT conservation $T^{\mu z}_{,z}=0$, so we must have 

\bea
T_L^{0z}&=&T_R^{0z}\nn
T_L^{zz}&=&T_R^{zz}\nn
T_L^{az}&=&T_R^{az}
\label{EMT}
\tea
$a=x,y$, where (L) refers to the half space $z<0$ and $R$ to $z>0$. The fluid is characterized by its temperature $T$ and its four velocity $u^{\mu}$ with $u^2=-1$, which may be further parameterized

\bea
u^0&=&\frac1{\sqrt{1-v^2}}\nn
u^z&=&\frac v{\sqrt{1-v^2}}\nn
u^a&=&0
\tea
The energy-momentum tensor has the ideal form for a conformal fluid (for simplicity we assume Maxwell-J\"uttner statistics)

\be
T^{\mu\nu}_{id}=\frac 1{\pi^2}T^4\left[4u^{\mu}u^{\nu}+\eta^{\mu\nu}\right],
\label{ideal}
\te
where $\eta^{\mu\nu}=\rm{diag}\;\left(-1,1,1,1\right)$ is the Minkowski metric. We then get

\bea
K&=&\frac4{\pi^2}T_L^4\frac {v_L}{1-v_L^2}=\frac4{\pi^2}T_R^4\frac {v_R}{1-v_R^2}\nn
K'&=&\frac1{\pi^2}T_L^4\frac{1+3v_L^2}{1-v_L^2}=\frac1{\pi^2}T_R^4\frac{1+3v_R^2}{1-v_R^2}
\label{shock}
\tea
$K$ is a constant which expresses the common value of $T^{0z}$ on both sides of the shock, similarly $K'$ represents the common value of $T^{zz}$. In the more complex theories to be considered below, temperature and velocity will no longer be constant on either side, but as long as energy-momentum is conserved, $T^{0z}$ and $T^{zz}$ will be constant, and $K$ and $K'$ will still represent them, respectively. Their actual value is defined by the asymptotic temperature and velocity, which we call $T_L$ and $v_L$ in all the models we shall consider.

Elliminating $T_{L,R}$ from eqs. (\ref{shock})

\be
3v_L^2-4Cv_L+1=0,
\te
where $C=K'/K$, so

\be
v_{L,R}=\frac2{3}\left\{C\pm \sqrt{C^2-\frac{3}4}\right\}
\te
There is a nontrivial shock when both roots are real and $\le 1$. In the allowed range we have

\be
v_Lv_R=\frac13
\label{soundspeed}
\te
We shall call $v_L$ the root such that $1/\sqrt 3\le v_L\le 1$, and then $1/\sqrt 3\ge v_R\ge 1/3$. Then

\be
\left(\frac {T_R}{T_L}\right)^4=\frac {v_L\left(1-v_R^2\right)}{v_R\left(1-v_L^2\right)}=\frac {\left(3v_L^2-\frac13\right)}{\left(1-v_L^2\right)}
\te
Observe that $T_R\ge T_L$ and so the entropy density behind the shock is greater than in front of it, in agreement with the Second Law.

\subsection{Shocks in Landau-Lifshitz theory}

A viscous fluid cannot sustain a discontinuity, but for a solution which depends only on $z$, integrating EMT conservation from $z=-\infty$ to $z=\infty$, we see that the relations (\ref{EMT}) hold for the asymptotic values. In particular, we may assume that $u^{x,y}\to 0$ asymptotically. We shall make the stronger assumption that the solution is axially symmetric around the $z$ direction everywhere. Thus we are seeking a solution depending only the $z$ coordinate and axially symmetric which asymptotically reduces to an ideal fluid when $z\to\pm\infty$, with boundary conditions obeing the junction conditions (\ref{EMT}) for an ideal fluid, namely conditions (\ref{shock}).

It is interesting to see the shock structure in Landau-Lifshitz theory, where

\be
T^{\mu\nu}=T^{\mu\nu}_{id}-\frac 1{\pi^2}\eta_0T^3\sigma^{\mu\nu}
\te
where $\eta_0$ is a dimensionless constant, essentially the viscosity to entropy ratio, and

\bea
\sigma^{\mu\nu}&=&\Delta^{\mu\rho}\Delta^{\nu\sigma}\left[u_{\rho,\sigma}+u_{\sigma,\rho}-\frac23\Delta_{\rho\sigma}u^{\lambda }_{,\lambda}\right]\nn
&=&\Delta^{\nu\sigma}u^{\mu}_{,\sigma}+\Delta^{\mu\rho}u^{\nu}_{,\rho}-\frac23\Delta^{\mu\nu}u^{\lambda}_{,\lambda}
\label{shear}
\tea

$\Delta^{\mu\rho}=\eta^{\mu\rho}+u^{\mu}u^{\rho}$. Since by definition $\sigma_{\mu\nu}u^{\mu}=0$, we must have

\bea
\sigma^{z0}&=&v\sigma^{zz}\nn
\sigma^{00}&=&v^2\sigma^{zz}
\tea
and

\be
\sigma^{zz}=\frac 43\frac{v_{,z}}{\left(1-v^2\right)^{5/2}}
\te

Now the constancy of $T^{0z}$ and $T^{zz}$ yields two equations

\bea
\frac 1{\pi^2}T^4\frac{4v}{1-v^2}-\frac 1{\pi^2}\eta_0T^3\frac 43\frac{vv_{,z}}{\left(1-v^2\right)^{5/2}}&=&K\nn
\frac 1{\pi^2}T^4\left[\frac{1+3v^2}{1-v^2}\right]-\frac 1{\pi^2}\eta_0T^3\frac 43\frac{v_{,z}}{\left(1-v^2\right)^{5/2}}&=&K'
\tea
with $K,K'=$ constant. With the boundary conditions that $v\to v_L$ and $T\to T_L$ as $z\to -{\infty}$ they are the same constants as in eq. (\ref{shock}); then $v\to v_R$ and $T\to T_R$ as $z\to {\infty}$. 
We may write

\be
\frac43\eta_0\frac{v_{,z}}{\left(1-v^2\right)^{3/2}}=-3T\frac{\left(v_L-v\right)\left(v-\frac1{3v_L}\right)}{\left[1-\frac v{4v_L}\left(3v_L^2+1\right)\right]}
\label{stop1}
\te
We see that $v_L$ and $v_R=1/\left(3v_L\right)$ are the only values of $v$ where $v_{,z}$ may vanish. $v$ goes monotonically from one to the other, reaching the limiting values only asymptotically. To find the speed of approach to the asymptotic value, we write $v=v_L-\vartheta$, with $\vartheta\propto e^{\lambda_{LL} z}$. Then to first orden in $\vartheta$ we get \cite{KKM10}

\be
\lambda_{LL}=3T_L\frac{\left(1-v_L^2\right)^{1/2}}{\eta_0v_L}\left(v_L^2-\frac1{3}\right)
\label{lambdall}
\te
After solving the equation for $v$, we may find the temperature from

\be
\frac{4vT^4}{1-v^2}\left[1+\frac34\frac{\left(v_L-v\right)\left(v-\frac1{3v_L}\right)}{1-\frac v{4v_L}\left(3v_L^2+1\right)}\right]=\frac{4v_LT_L^4}{1-v_L^2}
\te

\subsection{Israel-Stewart fluids}

In the Israel-Stewart or extended thermodynamics approach, the viscous part of $T^{\mu\nu}$ is left undetermined

\be
T^{\mu\nu}=T^{\mu\nu}_{id}+\Pi^{\mu\nu}
\te
Since $\Pi^{\mu\nu}u_{\nu}=0$, we have $\Pi^{0z}=v\Pi^{zz}$. Thus calling $\Pi^{zz}=\Pi$, we find

\bea
K=T^{0z}&=&\frac4{\pi^2}\frac{v_LT_L^4}{1-v_L^2}=\frac4{\pi^2}\frac{vT^4}{1-v^2}+v\Pi\nn
K'=T^{zz}&=&\frac3{\pi^2}\frac{T_L^4}{1-v_L^2}\left[v_L^2+\frac13\right]=\frac3{\pi^2}\frac{T^4}{1-v^2}\left[v^2+\frac13\right]+\Pi
\tea          '
Then

\be
\frac{\pi^2}3\left(1-v^2\right)T^{-4}\Pi=\frac{\left(v_L-v\right)\left(v-\frac1{3v_L}\right)}{\left[1-\frac {3v}{4v_L}\left(v_L^2+\frac13\right)\right]}
\label{ISpi}
\te
The system is closed by asking that $\Pi^{\mu\nu}$ relaxes to its Landau-Lifshitz form on a time-scale $\tau=\tau_0/T$

\be
\frac{\tau_0}Tu^{\rho}\Pi^{\mu\nu}_{,\rho}+\Pi^{\mu\nu}=-\frac 1{\pi^2}\eta_0T^3\sigma^{\mu\nu}
\label{relax}
\te
Taking the $zz$ component we get

\be
\frac{3T^4}{\pi^2}\frac{\left(v_L-v\right)\left(v-\frac1{3v_L}\right)}{\left(1-v^2\right)\left[1-\frac {3v}{4v_L}\left(v_L^2+\frac13\right)\right]}+\frac{\tau_0v}{T\sqrt{1-v^2}}\frac d{dz}\frac{3T^4}{\pi^2}\frac{\left(v_L-v\right)\left(v-\frac1{3v_L}\right)}{\left(1-v^2\right)\left[1-\frac {3v}{4v_L}\left(v_L^2+\frac13\right)\right]}=-\frac 4{3\pi^2}\frac{\eta_0T^3v_{,z}}{\left(1-v^2\right)^{5/2}}
\label{stop2}
\te
When $z\to -\infty$ we write $v=v_L-\vartheta e^{\lambda_{IS} z}$ and linearize on $\vartheta$

\be
\lambda_{IS}=3T_L\frac{\left(v_L^2-\frac1{3}\right)\left(1-v_L^2\right)^{1/2}} {v_L\left[\eta_0-3\tau_0\left(v_L^2-\frac13\right)\right]}
\label{lambdais}
\te
A suitable model must satisfy

\be
v_L^2-\frac13<\frac13\frac{\eta_0}{\tau_0}
\label{limit}
\te
Causality requires the right hand side to be strictly less than $2/3$ (see Appendixes \ref{CHEap} and \ref{signalprop}), and so this sets an upper bound for $v_L$ which is strictly less than $1$, otherwise there is no solution smoothly approaching equilibrium. For example, AdS-CFT yields a value $\eta_0=1$, $\tau_0=1-\left(\ln 2/2\right)$, and the criterion eq. (\ref{limit}) becomes $v_L^2\le 0.84$ \cite{KKM10}. Both the Chapman-Enskog and Grad approaches yield $\eta_0=\left(4/5\right)\tau_0$, and so the theory breaks down when $v_L^2\ge 3/5$.

\section{Kinetic theory}
\label{kt}

In this section we will show that, in kinetic theory under the relaxation time approximation, there are solutions smoothly approaching equilibrium regardless of the asymptotic velocity $v_L$. 

Under the relaxation time or Anderson-Witting approximation, the kinetic equation reads

\be
p^{\mu}f_{,\mu}=\frac T{\tau_0}\left(u_{eq\mu}p^{\mu}\right)\left[f-f_{eq}\right]
\label{AW}
\te

where $f_{eq}$ is the Maxwell-J\"uttner distribution with parameters $T_{eq}$, $u^{\mu}_{eq}$ defined by the consistency condition

\be
T^{\mu\nu}u_{eq\nu}=-\frac3{\pi^2}T_{eq}^4u^{\mu}_{eq}
\label{consisa}
\te
For simplicity, we shall call $T_{eq}=T$ and $v_{eq}=v$. Given the symmetries of the shock solution, the Boltzmann equation (\ref{AW}) reduces to 

\be
f_{,z}+\Lambda\left(z\right)f=\Lambda\left(z\right) f_{eq}
\te
Or else

\be
\left[f-f_{eq}\right]_{,z}+\Lambda\left[f-f_{eq}\right]=\phi_{,z}\;f_{eq}
\label{equation}
\te
where

\bea
\Lambda\left(z\right)&=&\frac{T\left[-u_{eq\mu}p^{\mu}\right]}{\left(p^z\tau_0\right)}\nn
f_{eq}&=&e^{-\phi\left(z\right)}\nn
\phi\left(z\right)&=&\frac{\left[-u_{eq\mu}p^{\mu}\right]}{T}
\tea
We are seeking the solution which reaches asymptotically equilibrium values as $z\to -\infty$. This is

\be
f\left(z\right)=f_{eq}\left(z\right)+\int_{-\infty}^zdz'\;e^{-\int_{z'}^zdz''\Lambda\left(z''\right)}\;\left[\phi_{,z}\;f_{eq}\right]\left(z'\right)
\label{solution}
\te
When we use this to compute the EMT, we find

\be
T^{\mu\nu}=T_{eq}^{\mu\nu}+t^{\mu\nu}
\te
The consistency condition (\ref{consisa}) becomes

\be
t^{\mu\nu}u_{eq\nu}=0
\label{consisb}
\te
This condition implies energy momentum tensor conservation, so we also get that $T^{0z}$ and $T^{zz}$ remain constant.

Let us analize $t^{\mu\nu}$ more closely:

\be
t^{\mu\nu}=\int_{-\infty}^zdz'\int Dp\;p^{\mu}p^{\nu}\;e^{-\int_{z'}^zdz''\Lambda\left(z''\right)}\;\left[\phi_{,z}\;f_{eq}\right]\left(z'\right)
\te
$Dp=2\delta\left(-p^2\right)\theta\left(p^0\right)d^4p/\left(2\pi\right)^3$ is the covariant momentum space volume element. For each fixed $z'$ introduce new variables 

\bea
q^0&=&\frac{p^0-v\left(z'\right)p^z}{\sqrt{1-v\left(z'\right)^2}}\nn
q^z&=&\frac{p^z-v\left(z'\right)p^0}{\sqrt{1-v\left(z'\right)^2}}\nn
q^{x,y}&=&p^{x,y}
\label{p2q}
\tea
($q_x^2=q_y^2=q_0^2-q_z^2=p_0^2-p_z^2$) then we may write

\be
\phi\left[p^{\mu},z'\right]=\frac{q^0}{T}
\label{dtt}
\te
Observe that we regard the transformation (\ref{p2q}) as a change of variables within the momentum integral at a given point in space, rather than as a global coordinate change. 

The $z$ derivative in eq. (\ref{solution}) is taken at $p$ held constant, so upon the change of variables we must write

\be
\left.\frac{\partial \phi}{\partial z}\right|_p=\left.\frac{\partial\phi}{\partial z}\right|_q+\frac{\partial\phi}{\partial q^{\mu}}\left.\frac{\partial q^{\mu}}{\partial z}\right|_p
\te
Actually the $q^{\mu}$ derivatives are non zero for $\mu=0$ and $z$ only, so

\be
\left.\frac{\partial \phi}{\partial z}\right|_p=\left.\frac{\partial \phi}{\partial z}\right|_q+\frac{\partial \phi}{\partial q^{0}}\left.\frac{\partial q^{0}}{\partial z}\right|_p+\frac{\partial\phi}{\partial q^{z}}\left.\frac{\partial q^{z}}{\partial z}\right|_p
\te
Now

\be
\left(\begin{array}{c} q^0 \\ q^z\end{array}\right)=\frac1{\sqrt{1-v^2}}\left(\begin{array}{cc} 1&-v \\ -v&1\end{array}\right)\left(\begin{array}{c} p^0 \\ p^z\end{array}\right)
\te

\be
\frac{\partial}{\partial z}\left(\begin{array}{c} q^0 \\ q^z\end{array}\right)=\frac{v'}{\left(1-v^2\right)}\left(\begin{array}{cc} 0&-1 \\ -1&0\end{array}\right)\left(\begin{array}{c} q^0 \\ q^z\end{array}\right)
\te

so
\be
\left.\frac{\partial \phi}{\partial z}\right|_p=\left.\frac{\partial \phi}{\partial z}\right|_q-\frac{v_{,z}}{\left(1-v^2\right)}\left[q^z\frac{\partial \phi}{\partial q^{0}}+q^0\frac{\partial \phi}{\partial q^{z}}\right]
\label{dtrans}
\te
Now the conditions (\ref{consisb}) become

\bea
0&=&\int_{-\infty}^z\frac{dz'}{T\left(z'\right)}\int Dq\left(q^0+v\left(z'\right)q^z\right)q^{0}e^{-\int_{z'}^zdz''\Lambda\left(z',z''\right)}\left[\frac{q^0T_{,z'}\left(z'\right)}{T\left(z'\right)}+\frac{q^zv_{,z'}\left(z'\right)}{\left(1-v^2\left(z'\right)\right)}\right]e^{-q^0/T\left(z'\right)}\nn
0&=&\int_{-\infty}^z\frac{dz'}{T\left(z'\right)}\int Dq\left(q^z+v\left(z'\right)q^0\right)q^{0}e^{-\int_{z'}^zdz''\Lambda\left(z',z''\right)}\left[\frac{q^0T_{,z'}\left(z'\right)}{T\left(z'\right)}+\frac{q^zv_{,z'}\left(z'\right)}{\left(1-v^2\left(z'\right)\right)}\right]e^{-q^0/T\left(z'\right)}
\tea
where

\be
\Lambda\left(z',z''\right)=\frac{T\left(z''\right)\left[\left(1-v\left(z'\right)v\left(z''\right)\right)q^0+\left(v\left(z'\right)-v\left(z''\right)q^z\right)\right]}{\left(q^z+v\left(z'\right)q^0\right)\tau_0\sqrt{1-v^2\left(z''\right)}}
\te
We define dimensionless momenta $q^{\mu}=T\left(z'\right)r^{\mu}$ and go to polar coordinates to get

\bea
0&=&\frac6{\pi^2}\int_{-\infty}^zdz'\;T^4\left(z'\right)\int_{-1}^1dx\;\left(1+v\left(z'\right)x\right)\;e^{-\int_{z'}^zdz''\Lambda\left(x; z',z''\right)}\;\left[\frac{T_{,z'}}{T\left(z'\right)}+\frac{xv_{,z'}\left(z'\right)}{\left(1-v^2\left(z'\right)\right)}\right]\nn
0&=&\frac6{\pi^2}\int_{-\infty}^zdz'\;T^4\left(z'\right)\int_{-1}^1dx\;\left(x+v\left(z'\right)\right)\;e^{-\int_{z'}^zdz''\Lambda\left(x; z',z''\right)}\;\left[\frac{T_{,z'}}{T\left(z'\right)}+\frac{xv_{,z'}\left(z'\right)}{\left(1-v^2\left(z'\right)\right)}\right]
\tea
where now

\be
\Lambda\left(x; z',z''\right)=\frac{T\left(z''\right)\left[1-v\left(z'\right)v\left(z''\right)+\left(v\left(z'\right)-v\left(z''\right)x\right)\right]}{\left(x+v\left(z'\right)\right)\tau_0\sqrt{1-v^2\left(z''\right)}}
\te
When $z\to-\infty$ we expect a solution where $T=T_L\left(1+te^{\lambda_{AW} z}\right)$, $v=v_L-\vartheta e^{\lambda_{AW} z}$. Linearizing we get

\bea
At-B\frac{\vartheta}{\left(1-v_L^2\right)}&=&0\nn
Ct-D\frac{\vartheta}{\left(1-v_L^2\right)}&=&0
\tea
where

\be
\kappa=v_L+\frac{T_L}{\lambda_{AW}\tau_0}\sqrt{1-v_L^2}
\label{param1}
\te
\bea
A&=&\int_{-1}^1dx\;\frac{\left(1+v_Lx\right)\left(x+v_L\right)}{x+\kappa}\nn
B&=&\int_{-1}^1dx\;x\frac{\left(1+v_Lx\right)\left(x+v_L\right)}{x+\kappa}\nn
C&=&\int_{-1}^1dx\;\frac{\left(x+v_L\right)^2}{x+\kappa}\nn
D&=&\int_{-1}^1dx\;x\frac{\left(x+v_L\right)^2}{x+\kappa}
\label{ABCD}
\tea
The dispersion relation $AD-BC=0$ reduces to (see Appendix \ref{param2der})
\be
0=v_L^2-\frac13-v_LG\left[\kappa\right]
\label{param2}
\te
where
\be
G\left[\kappa\right]=\kappa-\frac13\frac{J_0\left[\kappa\right]}{\kappa J_0\left[\kappa\right]-2}
\te
\be
J_0\left[k\right]=\ln\frac{k+1}{k-1}
\label{nonanal}
\te
Equations (\ref{param1}) and (\ref{param2}) define parametrically $\lambda_{AW}$ as a function of $v_L$.

Let's analyze the limiting cases. When $k\to\infty$, $G\left(k\right)\approx 4/\left(15k\right)$, so $v^2\to 1/3$,

\be
\kappa\approx\frac{4}{15}\frac{v_L}{v_L^2-\frac13}
\te

\be
\lambda_{AW}\approx\frac{T_L}{\tau_0}\frac{\sqrt{1-v_L^2}}{v_L\left(\frac35-v_L^2\right)}\left(v_L^2-\frac13\right)\approx \frac{15}4\sqrt{2}\frac{T_L}{\tau_0}\left(v_L^2-\frac13\right)
\label{asymp1}
\te
When $k\to 1$, $J_0\left(k\right)\approx- \ln\left(k-1\right)+\ln 2\to\infty$, $v_L\to 1$,

\be
G\left(\kappa\right)\approx \frac23\left(1-\frac1{\ln\left(k-1\right)}\right)
\te

\be
\kappa\approx 1+e^{-1/\left(2\left(1-v_L\right)\right)}
\te

\be
\lambda_{AW}\approx\frac{T_L}{\tau_0}\sqrt{\frac 2{1-v_L}}
\label{asymp2}
\te

\begin{center}
\begin{figure}[h]
\scalebox{0.8}{\includegraphics{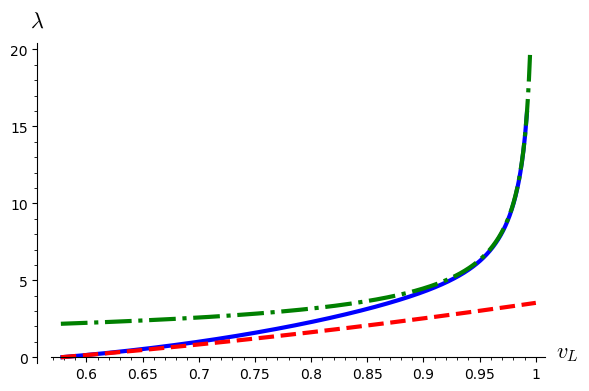}}
\caption{(Color online) Exact decay rate derived from kinetic theory with an Anderson-Witting collision term and its asymptotic forms: (blue, full line) exact decay rate, defined parametrically by eqs. (\ref{param1}) and (\ref{param2}); (red, dashes) asymptotic behavior for $v_L\to 1/\sqrt{3}$, eq. (\ref{asymp1}); (green, dots and dashes) asymptotic behavior for $v_L\to 1$, eq. (\ref{asymp2}). The divergence in the decay rate is stronger than predicted by holography, \cite{KKM10}. }
\label{f1plot}
\end{figure}
\end{center}

We can check that these analytical asymptotic forms match very well the exact solution eqs. (\ref{param1}) and (\ref{param2}) in their respective regimes, see fig. \ref{f1plot}.

We conclude that kinetic theory may describe the approach to equilibrium regardless of the limiting velocity, as long as $v_L<1$. The difference in behavior between kinetic theory and hydrodynamics may be traced back to the fact that, in kinetic theory, the speed of signal propagation is the maximum speed of the particles for which the distribution function is not zero \cite{I88}. For the near equilibrium distribution functions we are considering, that covers the full range, so always the asymptotic velocity $v_L$ shall be below the speed of signal propagation.

\section{Causal fluids}
\label{DTT}

We reduce kinetic theory to hydrodynamics by making the ansatz

\be
f=\exp\left\{\frac1{T}\left(u_{\mu}p^{\mu}\right)+\frac{\zeta_{\mu\nu}p^{\mu}p^{\nu}}{\left(-u_{\mu}p^{\mu}\right)}+\frac{\xi_{\mu\nu\rho}p^{\mu}p^{\nu}p^{\rho}}{\left(-u_{\mu}p^{\mu}\right)^2}\right\}
\te

for the distribution function. The tensors $\zeta_{\mu\nu}$ and $\xi_{\mu\nu\rho}$ are totally symmetric, transverse to $u^{\mu}$, and traceless on any pair of indexes. The equations for the coefficients are derived by taking moments of the kinetic equation, for which we assume the Anderson-Witting form (\ref{AW}); see Appendix \ref{entropy}.

Under our symmetry assumptions $u^{\mu}$ is characterized by the single velocity $v$ in the $z$ direction. Likewise, $\zeta_{\mu\nu}$ and $\xi_{\mu\nu\rho}$ contribute a single degree of freedom each. To see this, observe that in the local rest frame of the fluid, all components with a $0$ index must vanish, while symmetry implies that components with an odd number of $x,y$ components also vanish, $\zeta_{xx}=\zeta_{yy}$ and $\xi_{zxx}=\xi_{zyy}$. Now tracelessness implies that $\zeta_{xx}=\left(-1/2\right)\zeta_{zz}$ and $\xi_{zxx}=\left(-1/2\right)\xi_{zzz}$. Henceforth we shall call $\zeta$ and $\xi$ the single nontrivial component of these tensors in the local rest  frame. Introducing the momenta in the local rest frame as in eq. (\ref{p2q}) we may write

\be
f=\exp\left\{-\frac1{T}q^0+\zeta H_{\zeta}+\xi H_{\xi}\right\}
\label{dttloc}
\te
where

\bea
H_{\zeta}&=&\frac{\left(3q_z^2-q_0^2\right)}{2q^0}\nn
H_{\xi}&=&\frac{q^z\left(5q_z^2-3q_0^2\right)}{2q_0^2}
\label{haches}
\tea
We see that if $\xi\not=0$, $f$ is not even in $q^z$, and for this reason $v_{eq}\not=0$ in the local rest frame either. For example, let us consider again the definition of $T_{eq}$ and $u_{eq}$. After performing the change of variables (\ref{p2q}), which of course has unit Jacobian, we get, in the local rest frame of the fluid

\bea
\int\;Dq\;q^{0}\left(u_{eq}^0q^{0}-u_{eq}^zq^{z}\right)f&=&\frac 3{\pi^2}T_{eq}^4u_{eq}^{0}\nn
\int\;Dq\;q^{z}\left(u_{eq}^0q^{0}-u_{eq}^zq^{z}\right)f&=&\frac 3{\pi^2}T_{eq}^4u_{eq}^{z}\nn
\tea
or in terms of the velocity $v_{eq}$

\bea
\int\;Dq\;q^{0}\left(q^{0}-v_{eq}q^{z}\right)f&=&\frac 3{\pi^2}T_{eq}^4\nn
\int\;Dq\;q^{z}\left(q^{0}-v_{eq}q^{z}\right)f&=&\frac 3{\pi^2}T_{eq}^4v_{eq}
\tea
It is easy to see that $v_{eq}=0$ and $T_{eq}=T$ to first order in $\zeta$ and $\xi$, but not to higher order. This is related to the possibility of building vector fields out of $\zeta_{\mu\nu}$ and $\xi_{\mu\nu\rho}$, such as $\xi_{\mu\nu\rho}\zeta^{\nu\rho}$ or $\xi_{\mu\nu\rho}\xi^{\nu\lambda\sigma}\xi^{\rho}_{\lambda\sigma}$
 
\subsection{Energy momentum tensor}

Insofar as the energy momentum conservation conditions are still exact equations of the theory, and the energy momentum tensor in the rest frame of the shock depends only on $z$, we still have the identities

\bea
K=T^{0z}&=&\int\;Dp\;p^0p^z\;f=\rm{constant}\nn
K'=T^{zz}&=&\int\;Dp\;\left(p^z\right)^2\;f=\rm{constant}
\tea
With $K$ and $K'$ depending only on the asymptotic state as for an ideal fluid, see eq. (\ref{shock}).
Performing the change of variables (\ref{p2q}) and linearizing on $\zeta$ and $\xi$ we get

\bea
T^{0z}&=&\frac1{1-v^2}\int\;Dq\;\left(q^0+vq^z\right)\left(q^z+vq^0\right)\;f\nn
&=&\frac{4T^4}{\pi^2}\frac v{\left(1-v^2\right)}\left\{1 +\frac25\zeta T\right\}\nn
T^{zz}&=&\frac1{1-v^2}\int\;Dq\;\left(q^z+vq^0\right)^2\;f\nn
&=&\frac{3T^4}{\pi^2}\frac 1{\left(1-v^2\right)}\left\{v^2+\frac13+\frac{8}{15}\zeta T\right\}
\tea
This is equivalent to the Israel-Stewart energy momentum tensor identifying

\be
\Pi=\frac8{5\pi^2}\frac{T^4}{\left(1-v^2\right)}\zeta T
\te
We thus obtain two relations among $T$, $v$ and the dimensionless combination $\zeta T$. Elliminating $T$ we get

\be
\zeta T=\frac{15}8\frac{\left(v_L-v\right)\left(v-\frac1{3v_L}\right)}{\left[1-\frac34\frac{v}{v_L}\left(v_L^2+\frac13\right)\right]}
\label{stop3}
\te
which is equivalent to eq. (\ref{ISpi}), and further writing $v=v_L-\vartheta e^{\lambda_{DTT} z}$ and linearizing on $\vartheta$

\be
\zeta T=\frac52\frac{\left(v_L^2-\frac13\right)}{v_L\left(1-v_L^2\right)}\vartheta e^{\lambda_{DTT} z}
\label{z2v}
\te

\subsection{Equations of motion}

The equations of motion will have the form

\be
\int\;Dp\;H_{\alpha}\left(z,p\right)\;\left[p^{z}\left.\frac{\partial f}{\partial z}\right|_p-I_{col}\left[z,p\right]\right]=0
\te
for suitable functions $H_{\alpha}$ \cite{DMNR11,DMNR11b,DMNR12,DMNR12b}.

When we perform the transformation (\ref{p2q}) we must take into account that $z$ is not transformed (this is a change of variables, not a change of coordinates). So even if the function $H_{\alpha}$ is a particular component of a tensor, we do not transform it as such, but only as a given function of $z$ and $p$.
The same argument may be used to transform the collision integral, so

\be
\int\;Dq\;H_{\alpha}\left(z,q\right)\;\left[\frac{q^z+vq^0}{\sqrt{1-v^2}}\left.\frac{\partial f}{\partial z}\right|_p-I_{col}\left[z,q\right]\right]=0
\te

The $f$ derivative is transformed as in eq. (\ref{dtrans}).
It is convenient to move the derivatives out of the integral, observing that 

\be
\left.\frac{\partial}{\partial z}\delta\left(p^2\right)\right|_p=\left.\frac{\partial}{\partial z}p^z\right|_p=0
\te

in either the $-p$ or $-q$ representation. Therefore we get

\be
0=\frac d{dz}A_{\alpha}-B_{\alpha}-I_{\alpha}
\label{alphaeqs}
\te
where

\bea
A_{\alpha}&=&\int\;Dq\;H_{\alpha}\left(q\right)\;\frac{q^z+vq^0}{\sqrt{1-v^2}}f\nn
B_{\alpha}&=&\int\;Dq\;\left\{\left[\frac{\partial}{\partial z}-\frac{v'}{\left(1-v^2\right)}\left[q^z\frac{\partial }{\partial q^{0}}+q^0\frac{\partial }{\partial q^{z}}\right]\right]H_{\alpha}\right\}\frac{q^z+vq^0}{\sqrt{1-v^2}}f\nn
I_{\alpha}&=&\int\;Dq\;H_{\alpha}\left(q\right)\;I_{col}\left[z,q\right]
\tea
If we choose $H_0=p^0$ and $H_z=p^z$, $B_{0,z}$ and $I_{0,z}$ vanish and we obtain once again the constancy of the EMT.
Entropy considerations, discussed further in Appendix \ref{entropy}, suggest choosing the remaining functions as $H_{\zeta}$ and $H_{\xi}$ in equations (\ref{haches}). Then, linearizing on $\zeta$, $\xi$, and $v'$

\bea
A_{\zeta}&=&\frac{12}{5\pi^2}\frac {v_L\zeta}{\sqrt{1-v_L^2}}T_L^5+\frac{36}{35 \pi^2}\frac{\xi}{\sqrt{1-v_L^2}}T_L^5\nn
B_{\zeta}&=&-\frac{8}{5\pi^2}\frac{v'}{\left(1-v_L^2\right)^{3/2}}T_L^4\nn
I_{\zeta}&=&-\frac{12}{5\pi^2}\frac{\zeta}{\tau_0}\;T_L^6\nn
A_{\xi}&=&\frac{36}{35 \pi^2}\frac{\zeta T_L^5}{\sqrt{1-v_L^2}}+\frac{12}{7 \pi^2}\frac{v_L\xi T_L^5}{\sqrt{1-v_L^2}}\nn
B_{\xi}&=&0\nn
I_{\xi}&=&-\frac{12}{7 \pi^2}\frac{\xi T_L^6}{\tau_0}
\label{stop4}
\tea

If we do not include the $\xi$ term, we revert to the equations derived from Grad's ansatz, identifying $C^{\mu\nu}=\zeta^{\mu\nu}$.

Assuming that all variables depend on $z$ as $e^{\lambda_{DTT} z}$, the set of equations (\ref{z2v}, \ref{alphaeqs}) becomes
\be
\left(\begin{array}{ccc}
-\frac23\lambda_{DTT} & \lambda_{DTT} v_L+\frac {T_L}{\tau_0}\sqrt{1-v_L^2}&\frac37\lambda_{DTT}\\
v_L^2-\frac13& -\frac25v_L&0\\
0&\frac35\lambda_{DTT}&\lambda_{DTT} v_L+\frac {T_L}{\tau_0}\sqrt{1-v_L^2}\end{array}\right)\left(\begin{array}{c}\frac{\vartheta}{{1-v_L^2}}\\ \zeta T\\ \xi T\end{array}\right)=0
\te
writing

\be
\lambda_{DTT}=\frac{\alpha T_L}{v_L\tau_0}\sqrt{1-v_L^2}
\label{lambdadtt}
\te

The allowed values of $\alpha$ are the roots of

\be
\frac{\alpha^2}{v_L^2}\left[\frac67v_L^2-v_L^4-\frac3{35}\right]+2\alpha\left[\frac7{15}-v_L^2\right]+\frac13-v_L^2=0
\label{alpha}
\te

There will be a positive root as long as the coefficient of $\alpha^2$ is positive, which ceases to be true when $v_L^2\approx 0.74$; see Appendix \ref{signalprop}.

\section{Results and final remarks}
\label{results}

\begin{center}
\begin{figure}[h]
\scalebox{0.8}{\includegraphics{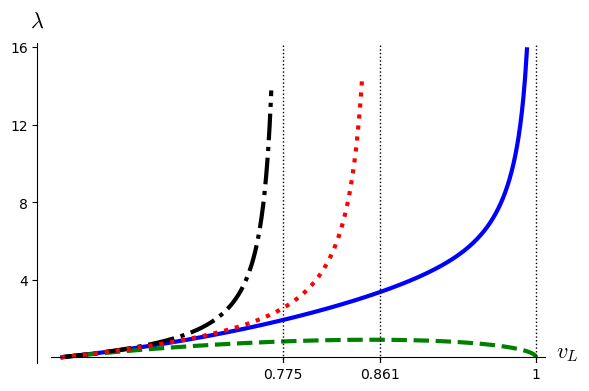}}
\caption{(Color online) The decay rates for the different theories discussed in this work: (blue, full line) exact decay rate, defined parametrically by eqs. (\ref{param1}) and (\ref{param2}); (green, dashes) the decay rate from Landau-Lifshitz theory, eq. (\ref{lambdall}); (black, dots and dashes) decay rate for an Israel-Stewart fluid with the constitutive relation derived from Grad's ansatz, $\eta_0=4\tau_0/5$, eq. (\ref{lambdais}); (red, dots) the decay rate derived from the DTT including the third momentum of the distribution function, eqs. (\ref{lambdadtt}) and (\ref{alpha}). The vertical grid lines show the characteristic speeds of the Israel-Stewart and DTT models, see Appendix \ref{signalprop}.}
\label{f2plot}
\end{figure}
\end{center}

In this paper we have computed the decay rate of the solution toward equilibrium at velocity $v_L$ as $z\to -\infty$ for several models, namely ideal fluids (where there is no decay), Landau-Lifshitz (eq. (\ref{lambdall})), Israel-Stewart (eq. (\ref{lambdais})), which actually holds for any theory which reduces to the Grad ansatz in the linear regime, kinetic theory with a relaxation time or Anderson-Witting collision term (eqs. (\ref{param1}) and (\ref{param2})), and finally for a DTT including third moments of the distribution function (eqs. (\ref{lambdadtt}) and (\ref{alpha})). The results are summarized in fig. (\ref{f2plot}).

The ``exact'' calculation yields a decay rate which ressembles the one derived from AdS-CFT correspondence \cite{KKM10} but with a stronger divergence in the upper limit; it diverges as $\gamma=\left(1-v_L^2\right)^{-1/2}$ while the result from holography diverges as $\gamma^{1/2}$ \cite{KKM10}.

Landau-Lifshitz provides a regular solution for any $v_L$, but the quantitative agreement to the ``exact'' result is not satisfactory beyond weak shocks.

Both the Israel-Stewart and DTT decay rates blow up at a finite value of $v_L$ set up by the highest speed of signal propagation (as we show in Appendix \ref{signalprop}). This exercise therefore provides a concrete example of the scenario discussed in \cite{OH90} and \cite{JP91}.

It is remarkable that if we extended the kinetic theory analysis to complex values of the asymptotic velocity $v_L$, then the decay rate $\lambda$ would be an analytic function of $v_L$ with a cut in the complex plane, signaled by the appearance of a logarithm in eq. (\ref{nonanal}). These non analiticities are a generic feature of kinetic theories that very much define the limit of validity of hydrodynamics \cite{KW19}.

If the limiting factor for hydrodynamics is that it cannot handle fast asymptotic velocities, then there should be no problem in the right hand side of the shock, where velocities are subsonic throughout. The decay rate can be computed with a straightforward adaptation of the arguments above (see Appendix \ref{rightside}), we show the result in fig. (\ref{f3plot}). As expected, there is no divergence in any of the models we are considering, but again the DTT outperforms the Landau-Lifshitz and Israel-Stewart schemes as a quantitative match to kinetic theory.

The fact that the speed of signal propagation sets the upper asymptotic velocity for which a regular solution exists may be easier to understand if we regard the time-independent configurations we have analyzed in this paper as the long time limit of the actual process by which the shock is formed. Remember that we are describing the fluid in the frame where the shock is at rest and the fluid advances from the left at velocity $v_L$. We may as well use the frame where the fluid is at rest and the shock advances to the left at velocity $-v_L$. Now picture the shock as a piston which materializes at $t=0$ at the position $z=0$, and then starts moving against fluid at rest. If $v_L<c$ the speed of signal propagation (see Appendix \ref{signalprop}), then the influence of the piston will outrun the piston itself. At time $t$, the fluid will remain at rest for all $z\le -ct$, and there will be a buffer zone between $z=-ct$ and the piston at $z=-v_Lt$. At long times and finite distances from the piston, in the frame where the piston is at rest, the fluid in the buffer zone will settle to a steady flow; this is the configuration we have described in this paper. However, if $v_L\ge c$ this is not possible; the piston keeps pushing against fluid at rest, and the hydrodynamic solution, if it exists at all, must be discontinuous \cite{OH90}.

\begin{center}
\begin{figure}[h]
\scalebox{0.8}{\includegraphics{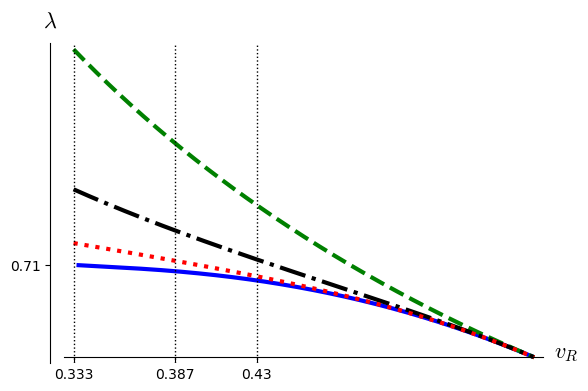}}
\caption{(Color online) The decay rates for the different theories discussed in this work on the right hand side of the shock: (blue, full line) exact decay rate, defined parametrically by eqs. (\ref{param3R}); (green, dashes) the decay rate from Landau-Lifshitz theory, eq. (\ref{lambdallR}); (black, dots and dashes) decay rate for an Israel-Stewart fluid with the constitutive relation derived from Grad's ansatz, $\eta_0=4\tau_0/5$, eq. (\ref{lambdaisR}); (red, dots) the decay rate derived from the DTT including the third momentum of the distribution function, eqs. (\ref{lambdadttR}) and (\ref{alphaR}). The vertical grid lines show the asymptotic right side velocities corresponding to the characteristic speeds of the Israel-Stewart and DTT models, see Appendix \ref{signalprop}, which mark their applicability limit.}
\label{f3plot}
\end{figure}
\end{center}

In spite of its limitations, figs. (\ref{f2plot}) and (\ref{f3plot}) show that including third moments in the DTT allows for a much more accurate description of the convergence to equilibrium. This should be considered along with the results of \cite{PC21} in choosing the correct hydrodynamic framework for a concrete application.

\section*{Acknowledgments}
I thank A. Kandus, N. Mir\'on Granese, L. Cantarutti, M. Nigro, G. E. Perna, J. Ruffinelli, Jer\'onimo Peralta Ramos and Leonardo Leitao for multiple discussions.
This work  was supported in part by Universidad de Buenos Aires through
grant UBACYT 20020170100129BA, CONICET and ANPCyT.

\appendix

\section{Chapman-Enskog and Grad}
\label{CHEap}

The Chapman-Enskog and Grad approaches attempt to anchor hydrodynamics on kinetic theory. 

Under the Chapman-Enskog approach, we seek a solution of the kinetic equation (\ref{AW}) of the form

\be
f=e^{u_{\mu}p^{\mu}/T}\left[1+\delta f\right]
\label{decomp}
\te
Then, see Appendix \ref{CHEder},
\be
\delta f=-\frac{\tau_0}{2T^2\left|u_{\rho}p^{\rho}\right|}p^{\mu}p^{\nu}\sigma_{\mu\nu}
\label{CHE}
\te
where $\sigma_{\mu\nu}$ is defined in eq. (\ref{shear}), leading to

\be
\Pi^{\mu\nu}=-\frac4{5\pi^2}\tau_0 T^3\sigma^{\mu\nu}
\label{CHEPI}
\te
which is the Landau-Lifshiz ansatz under the identification

\be
\eta_0=\frac45\tau_0
\label{CHEeta}
\te
If we use this value of $\eta_0$ in the equations from the Israel-Stewart approach, we find the theory becomes singular when $v_L^2=3/5$.

In the Grad approach, we write a decomposition (\ref{decomp}) but with a less constrained perturbation

\be
\delta f=\frac{p^{\mu}p^{\nu}}{\left|u_{\rho}p^{\rho}\right|}C_{\mu\nu}
\label{Graddf}
\te
where $u^{\mu}C_{\mu\nu}=C^{\mu}_{\mu}=0$. This satisfies the constraints (\ref{constraints}) and leads to 

\be
\Pi^{\mu\nu}=\frac8{5\pi^2} T^5C^{\mu\nu}
\label{GradPI}
\te
(compare to eq. (\ref{CHEPI})). To determine $C^{\mu\nu}$ we ask that some second moment of the Boltzmann equation is satisfied, or, using the linearity of the kinetic equation, simply substitute eq. (\ref{Graddf}) into eq. (\ref{AW}) \cite{DMNR10}, getting

\be
u^{\rho}C^{\mu\nu}_{,\rho}+\frac T{\tau_0}C^{\mu\nu}+\frac1{2T}\sigma^{\mu\nu}=0
\te
We may use eq. (\ref{GradPI}) to transform this to an equation for $\Pi^{\mu\nu}$, which turns out to be eq. (\ref{relax}) with the same $\tau_0$ and $\eta_0$ given by eq. (\ref{CHEeta}). As we already know, this leads to a theory breakdown when $v_L^2\ge 3/5$, see eq.  (\ref{limit}).

\section{Derivation of eq. (\ref{CHE})}
\label{CHEder}

To make the decomposition (\ref{decomp}) unique, we assume the constraints

\be
\int\;Dp\;p^{\mu}\left(-u_{\nu}p^{\nu}\right)e^{u_{\mu}p^{\mu}/T}\delta f=0
\label{constraints}
\te
This means the $e^{u_{\mu}p^{\mu}/T}=f_{eq}$. Then, assuming that the derivatives of $\beta^{\mu}/T$ are ``small'', we solve eq. (\ref{AW}) to first order to get

\be
\delta f=-\frac{\tau_0}{T\left|u_{\rho}p^{\rho}\right|}p^{\mu}p^{\nu}\beta_{\mu,\nu}
\label{CHEb}
\te
The constraints (\ref{constraints}) become the ideal hydrodynamic equations

\bea
\frac{\dot T}T+\frac13u^{\lambda}_{,\lambda }&=&0\nn
\dot u^{\mu}+\Delta^{\mu\nu}\frac{T_{,\nu}}T&=&0
\label{idealeqs}
\tea
For fields $T$ and $u^{\mu}$ satisfying eqs. (\ref{ideal}) we may simplify

\be
\beta_{\mu,\nu}+\beta_{\nu,\mu}=\frac1{T}\left[u_{\mu,\nu}+u_{\nu,\mu}+u_{\mu}\dot u_{,\nu}+u_{\nu}\dot u_{,\mu}-\frac23u_{\mu}u_{\nu}u^{\lambda }_{,\lambda}\right]
\te
and subtracting a term proportional to $\eta_{\mu\nu}$, which does not contribute to $f$ because $p^2=0$, we may substitute

\be
\beta_{\mu,\nu}+\beta_{\nu,\mu}\to \frac1T\sigma_{\mu\nu},
\te
where $\sigma_{\mu\nu}$ is the shear tensor (\ref{shear}), in eq. (\ref{CHEb}).

\section{Derivation of eq. (\ref{param2})}
\label{param2der}

Call 

\be
J_n=\int_{-1}^1dx\;\frac{x^n}{x+\kappa}
\te
Then the functions $A$, $B$, $C$ and $D$ from eq. (\ref{ABCD})
\bea
A&=&J_1+v_L\left(J_0+J_2\right)+v_L^2J_1\nn
B&=&J_2+v_L\left(J_1+J_3\right)+v_L^2J_2\nn
C&=&J_2+2v_LJ_1+v_L^2J_0\nn
D&=&J_3+2v_LJ_2+v_L^2J_1
\tea
and the dispersion relation is

\bea
0&=&v_L^4J_1^2+v_L^3J_1\left(J_0+3J_2\right)+v_L^2\left(J_1\left(J_1+J_3\right)+2J_2\left(J_0+J_2\right)\right)+v_L\left(J_3\left(J_0+J_2\right)+2J_1J_2\right)+J_1J_3\nn
&-&\left[v_L^4J_0J_2+v_L^3\left(2J_1J_2+J_0\left(J_1+J_3\right)\right)+v_L^2\left(J_2^2+J_0J_2+2J_1\left(J_1+J_3\right)\right)+v_LJ_2\left(3J_1+J_3\right)+J_2^2\right]
\tea
or else

\be
0=\left(1-v_L^2\right)\left(v_L^2\left(J_0J_2-J_1^2\right)+v_L\left(J_0J_3-J_1J_2\right)+\left(J_1J_3-J_2^2\right)\right)
\te
$J_0$ is defined in eq. (\ref{nonanal}). The remaining $J$ functions obey the recursion relations

\bea
J_1&=&2-\kappa J_0\nn
J_2&=&-\kappa J_1=\kappa^2J_0-2\kappa\nn
J_3&=&\frac23-\kappa J_2=\frac23+2\kappa^2-\kappa^3J_0
\tea
so we get

\be
0=v_L^2\left(\kappa J_0-2\right)+v_L\left(\frac13J_0-\kappa^2 J_0+2\kappa\right)+\frac13\left(2-\kappa J_0\right)
\te
which yields eq. (\ref{param2}) immediately.

\section{Entropy and the equations of motion}
\label{entropy}

Recall the entropy flux from kinetic theory \cite{RZ13,I88}

\bea
S^{\mu}&=&\int\;Dp\;p^{\mu}f\;\left[1-\ln f\right]\nn
&=&\Phi^{\mu}-\beta_{\nu}T^{\mu\nu}-\zeta_{\nu\rho}A^{\mu\nu\rho}
\tea
Where the Massieu function current

\be
\Phi^{\mu}=\int\;Dp\;p^{\mu}f
\te
is the potential for the hydrodynamic tensors, for example

\be
T^{\mu\nu}=\frac{\partial \Phi^{\mu}}{\partial \beta_{\nu}}
\te
We have made use of the symmetry of the shock wave problem to reduce the number of unknowns to just scalar variables. Moreover, we have seen that we may write $f=exp\left(-\phi\right)$. To set up the hydrodynamic formulation, we assume $\phi$ is an homogeneous function of the rest frame momenta $q^{\mu}$ of degree one, namely

\be
\phi=\frac{q^0}{T}\varphi\left[z,x\right]
\te
where $x=\cos\theta=q^z/q^0$. The function $\varphi$ may be expanded in Legendre polynomials of the variable $x$ \cite{DMNR12b}

\be
\varphi=1-\sum_{\ell\ge 1}Z_{\ell}\left[z\right]P_{\ell}\left(x\right)
\te
We move from kinetic theory to hydrodinamics when we truncate this series \cite{DMNR11,DMNR11b,DMNR12,DMNR12b}: ideal hydrodynamics keeps only $\ell=0$ and $1$, but assumes that $Z_1=0$ in the local rest frame; Israel-Stewart keeps $\ell=0,1$ and $2$, but linearizes on $Z_2$, once again forcing $Z_1=0$. The DTT presented above keeps $Z_0=1$, $Z_2=\zeta T$ and $Z_3=\xi T$ (up to normalization of the Legendre polynomials), with $Z_1=0$ to linear order in $Z_2$ and $Z_3$.  We introduce a dimensionless momentum $r^{\mu}=q^{\mu}/T$. The relevant components of $T^{\mu\nu}$ are

\bea
K=T^{0z}&=&\frac{T^4}{1-v^2}\int\;Dr\;\left(r^0+vr^z\right)\left(r^z+vr^0\right)\;e^{-r^0\left(1-\sum_{\ell}Z_{\ell}P_{\ell}\right)}\nn
K'=T^{zz}&=&\frac{T^4}{1-v^2}\int\;Dr\;\left(r^z+vr^0\right)^2\;e^{-r^0\left(1-\sum_{\ell}Z_{\ell}P_{\ell}\right)}
\tea
The Second Law reads

\be
S^z_{,z}=\sigma\ge 0
\te
so we only need the $z$ component of the entropy flux

\be
S=S^z=\frac{T^3}{\sqrt{1-v^2}}\int\;Dr \left(r^z+vr^0\right)\left[1+r^0\left(1-\sum_{\ell}Z_{\ell}P_{\ell}\right)\right]e^{-r^0\left(1-\sum_{\ell}Z_{\ell}P_{\ell}\right)}
\te
write

\be
r^0=\frac{\left(r^0+vr^z\right)-v\left(r^z+vr^0\right)}{1-v^2}
\te
to get

\be
S=\Phi+\frac1{\sqrt{1-v^2}T}\left[T^{0z}-vT^{zz}\right]-\frac1{T}\sum_{\ell}Z_{\ell}A_{\ell}
\te
where

\bea
\Phi&=&\frac{T^3}{\sqrt{1-v^2}}\int\;Dr \left(r^z+vr^0\right)e^{-r^0\left(1-\sum_{\ell}Z_{\ell}P_{\ell}\right)}\nn
A_{\ell}&=&\frac{T^4}{\sqrt{1-v^2}}\int\;Dr \left(r^z+vr^0\right)r^0P_{\ell}\left[x\right]\;e^{-r^0\left(1-\sum_{\ell}Z_{\ell}P_{\ell}\right)}
\tea
Then

\bea
S'&=&\frac{T'}T\left[3\Phi-\frac1{\sqrt{1-v^2}T}\left[T^{0z}-vT^{zz}\right]+\frac1{T}\sum_{\ell}Z_{\ell}A_{\ell}\right]\nn
&+&v'\left[\frac{\partial\Phi}{\partial v}+\frac{vT^{0z}-T^{zz}}{\left(1-v^2\right)^{3/2}T}\right]+\sum_{\ell}\left[Z'_{\ell}\left[\frac{\partial\Phi}{\partial Z_{\ell}}-\frac1{T}A_{\ell}\right]-\frac1{T}Z_{\ell}A'_{\ell}\right]
\tea
It is clear that the coefficient of $Z'_{\ell}$ vanishes. Now compute

\be
\frac{\partial\Phi}{\partial v}=\frac{T^3}{\left(1-v^2\right)^{3/2}}\int\;\frac{d^3r}{\left(2\pi\right)^3r^0} \left(vr^z+r^0\right)e^{-r^0\left(1-\sum_{\ell}Z_{\ell}P_{\ell}\right)}
\te
on the other hand

\be
\frac1T\left[vT^{0z}-T^{zz}\right]=-T^3\int\;\frac{d^3r}{\left(2\pi\right)^3r^0}r^z\left(r^z+vr^0\right)e^{-r^0\left(1-\sum_{\ell}Z_{\ell}P_{\ell}\right)}
\te
but

\be
-\frac{r^z}{r^0}e^{-r^0}=\frac{\partial}{\partial r^z}e^{-r^0}
\te
so integrating by parts

\be
\frac1T\left[vT^{0z}-T^{zz}\right]=-T^3\int\;\frac{d^3r}{\left(2\pi\right)^3r^0}\left(r^0+vr^z+\left(r^z+vr^0\right)\sum_{\ell}Z_{\ell}r^0\frac{\partial}{\partial r^z}\left[r^0P_{\ell}\right]\right)e^{-r^0\left(1-\sum_{\ell}Z_{\ell}P_{\ell}\right)}
\te
Where we are regarding $r^0$ as $\sqrt{r^{z2}+r^{x^2}+r^{y2}}$ rather than as the independent $0$ component of the $r^{\mu}$ vector; the relationship among the two approaches is

\be
r^0\frac{\partial }{\partial r^z}=r^0\left.\frac{\partial }{\partial r^z}\right|_{r^0}+r^z\left.\frac{\partial }{\partial r^0}\right|_{r^z}
\te
Next consider

\bea
&&\frac1{\sqrt{1-v^2}T}\left[T^{0z}-vT^{zz}\right]=\frac{T^3}{\sqrt{1-v^2}}\int\;\frac{d^3r}{\left(2\pi\right)^3r^0}\left(r^z+vr^0\right)r^0e^{-r^0\left(1-\sum_{\ell}Z_{\ell}P_{\ell}\right)}\nn
&=&\frac{T^3}{\sqrt{1-v^2}}\int\;\frac{d^3r}{\left(2\pi\right)^3r^0}\left(r^z+vr^0\right)e^{r^0\sum_{\ell}Z_{\ell}P_{\ell}}\left(-\vec{r}\cdot\nabla\right)e^{-r^0}\nn
&=&3\Phi+\frac{T^3}{\sqrt{1-v^2}}\int\;\frac{d^3r}{\left(2\pi\right)^3r^0}\left(r^z+vr^0\right)e^{-r^0\left(1-\sum_{\ell}Z_{\ell}P_{\ell}\right)}\sum_{\ell}Z_{\ell}\left(\vec{r}\cdot\nabla\right)r^0P_{\ell}\nn
\tea
If $H$ is a homogeneous function of degree $n$, then $\left(\vec{r}\cdot\nabla\right)H=nH$. In our case $n=1$, and we get

\be
\frac1{\sqrt{1-v^2}T}\left[T^{0z}-vT^{zz}\right]=3\Phi+\frac1{T}\sum_{\ell}Z_{\ell}A_{\ell}
\te
Therefore

\be
S'=-\frac{1}{T}\sum_{\ell}Z_{\ell}\left[A'_{\ell}-B_{\ell}\right]
\te
where

\be
B_{\ell}=-\frac{v'T^4}{{\left(1-v^2\right)^{3/2}}}\int\;\frac{d^3r}{\left(2\pi\right)^3r^0}\left(r^z+vr^0\right)\left(r^0\frac{\partial r^0P_{\ell}}{\partial r^z}\right)e^{-r^0\left(1-\sum_{\ell}Z_{\ell}P_{\ell}\right)}
\te
which agrees with our result above. 
To enforce the second law we need equations of motion of the form

\be
A'_{\ell}-B_{\ell}=I_{\ell}
\te
such that $\sum_{\ell}Z_{\ell}I_{\ell}\le 0$. The natural choice is

\be
I_{\ell}=\int\;Dq\;q^0P_{\ell}\;I_{col}
\te
since then the nonpositivity is enforced by the $H$ theorem.

\section{DTT characteristics}
\label{signalprop}

We shall investigate the characteristics of a DTT. We need to reinstate the time-dependence, but we shall only consider linearized deviations from rest. The equations are

\bea
\dot T^{00}+T^{0z}_{,z}&=&0\nn
\dot T^{z0}+T^{zz}_{,z}&=&0\nn
\dot A^0_{\zeta}+A^z_{\zeta,z}-B^0_{\zeta}\frac{\dot v}{\left(1-v^2\right)}-B^z_{\zeta}\frac{v_{,z}}{\left(1-v^2\right)}&=&I_{\zeta}\nn
\dot A^0_{\xi}+A^z_{\xi,z}-B^0_{\xi}\frac{\dot v}{\left(1-v^2\right)}-B^z_{\xi}\frac{v_{,z}}{\left(1-v^2\right)}&=&I_{\xi}
\tea
$A^z_{\zeta}$, $B^z_{\zeta}$, $A^z_{\xi }$ and $B^z_{\xi}$ have already been computed in the main text (where we ommitted the $z$ superscript), same as $I_{\zeta}$ and $I_{\xi}$. We introduce dimensionless variables $Z=\zeta T$ and $X=\xi T$, and further write $T=T_0e^t$, where $t$ is the linear deviation from equilibrium. Then

\bea
T^{00}&=&\frac{T_0^4}{\pi^2}3\left(1+4t\right);\;\;T^{0z}=\frac{T_0^4}{\pi^2}4v;\;\;T^{zz}=\frac{T_0^4}{\pi^2}\left[1+4t+\frac85Z\right]\nn
A^0_{\zeta}&=&\frac{T_0^4}{\pi^2}\frac{12}5Z;\;\;
A^z_{\zeta}=\frac{T_0^4}{\pi^2}\frac{36}{35}X;\;\;
B^0_{\zeta}=0;\;\;
B^z_{\zeta}=\frac{T_0^4}{\pi^2}\frac85;\;\;
I_{\zeta}=-\frac{T_0^5}{\pi^2}\frac1{\tau_0}\frac{12}5Z\nn
A^0_{\xi}&=&\frac{T_0^4}{\pi^2}\frac{12}7X;\;\;
A^z_{\xi}=\frac{T_0^4}{\pi^2}\frac{36}{35}Z;\;\;
B^0_{\xi}=B^z_{\xi}=0;\;\;
I_{\xi}=-\frac{T_0^5}{\pi^2}\frac1{\tau_0}\frac{12}7X
\tea
If we call $X^{a}=\left(t,v,Z,X\right)$, we get equations of the form $\dot X^a+\Gamma^a_bX'^b+\Lambda^a_bX^b=0$. We are interested on the penetration of a front into fluid at rest. The variables $X^a=0$ at the front and are continuous across the front, but the first derivatives $X'^a$ are not. Since the $X^a$ remain constant as we move along with the front with speed $c$, at the front $\dot X^a+cX'^a=0$. From the equations of motion this means that $\left[\Gamma^a_b-c\delta^a_b\right]X'^b=0$. We thereby get the dispersion relation as

\be
{\rm{det}}\;\left(\begin{array}{cccc}c&-\frac13&0&0\\-1&c&-\frac25&0\\0&-\frac23&c&-\frac37\\0&0&-\frac35&c\end{array}\right)=0
\te
Not including either $Z$ or $X$ is equivalent to considering only the upper left $2\times 2$ block; we thus get the usual result $c^2=1/3$. For any $v_L>c$ the ideal fluid solution is discontinuous.

Including $Z$ but not $X$ means considering only the upper left $3\times 3$ block. We then get $c^2=3/5$, which we recognize as the upper value of $v_L$ for an Israel-Stewart fluid under the constitutive relation $\eta_0=4\tau_0/5$ as demanded by the Grad approximation.

Finally, the characteristic velocity for the full theory is

\be
c^4-\frac67c^2+\frac3{35}=0
\te
with roots

\be
c^2=\frac37\left[1+\sqrt{\frac8{15}}\right]\approx 0.74
\te
and $c'^2=3/\left(35c\right)\approx 0.11$. We recognize that the coefficient of the leading term in eq. (\ref{alpha}) may be written as

\be
\frac67v_L^2-v_L^4-\frac3{35}=\left(c^2-v_L^2\right)\left(v_L^2-c'^2\right)
\te
and since $v_L\ge 1/\sqrt{3}>c'$, positivity of this coefficient, and therefore existence of a solution, requires $v_L\le c$.

A similar calculation yields the characteristic speed of an Israel-Stewart model. Linearizing the equations of motion around a static equilibrium ($v=\Pi=0$) we get

\bea
3\frac{\dot T}T+v'&=&0\nn
\frac{T'}T+\dot v+\frac{\pi^2}{4T^4}\Pi'&=&0\nn
\frac{\eta_0}{\tau_0}v'+\frac{\pi^2}{4T^4}\dot\Pi&=&0
\tea
On the front we have $\dot v=-Vv'$ and likewise for $T$ and $\Pi$, so we get the characteristic velocities as $V=0$ or 

\be
V^2=\frac13\left[1+\frac{\eta_0}{\tau_0}\right]
\te
which shows that the Israel-Stewart model must break down at a finite velocity, see eq. (\ref{limit}).

\section{The right hand side of the shock}
\label{rightside}

If the drawback of  hydrodynamics is not being able to handle fast asymptotic velocities, then the approach to equilibrium on the right side, where speeds are subsonic throughout, should pose no problems.

Let us start with Landau-Lifshitz fluids. Eq. (\ref{stop1}) is still valid, observe that we can write indistinctingly $v_L$ or $v_R=1/3v_L$. Now we write $v=v_R+\vartheta e^{-\lambda_{LL}z}$ and linearize, getting the equivalent to eq. (\ref{lambdall})

\be
\lambda^R_{LL}=T_R\frac{\left(1-v_R^2\right)^{1/2}}{\eta_0v_R}\left(1-3v_R^2\right)
\label{lambdallR}
\te
We now move to Israel-Stewart fluids. Up to eq. (\ref{stop2}) nothing changes, then we write $v=v_R+\vartheta e^{-\lambda_{IS}z}$ and linearize, getting, instead of eq. (\ref{lambdais})

\be
\lambda^R_{IS}=T_R\frac{\left(1-3v_R^2\right)\left(1-v_R^2\right)^{1/2}} {v_R\left[\eta_0+\tau_0\left(1-3v_R^2\right)\right]}
\label{lambdaisR}
\te
where we further set $\eta_0=4/5\;\tau_0$ as derived from the Grad approximation.

In kinetic theory, the solution that goes to equilibrium as $z\to\infty$ is (cfr. eq. (\ref{solution}))

\be
f\left(z\right)=f_{eq}\left(z\right)-\int_z^{\infty}dz'\;e^{\int_z^{z'}dz''\Lambda\left(z''\right)}\;\left[\phi_{,z}\;f_{eq}\right]\left(z'\right)
\label{Rsolution}
\te
Once again we find $T^{\mu\nu}=T^{\mu\nu}_{id}-t^{\mu\nu}$ with $t^{\mu\nu}u_{eq\nu}=0$. For large $z$, $T=T_R\left(1-te^{-\lambda z}\right)$ and $v=v_R+\vartheta e^{-\lambda z}$. The analysis carries on as in the text, and we get the dispersion relations

\bea
A_Rt-B_R\frac{\vartheta}{\left(1-v_R^2\right)}&=&0\nn
C_Rt-D_R\frac{\vartheta}{\left(1-v_R^2\right)}&=&0
\tea
where (cfr. eq. (\ref{param1}) and (\ref{ABCD}))

\be
\kappa_R=\frac{T_R}{\lambda^R_{AW}\tau_0}\sqrt{1-v_R^2}-v_R
\label{param1R}
\te
\bea
A_R&=&\int_{-1}^1dx\;\frac{\left(1+v_Rx\right)\left(x+v_R\right)}{x-\kappa_R}\nn
B_R&=&\int_{-1}^1dx\;x\frac{\left(1+v_Rx\right)\left(x+v_R\right)}{x-\kappa_R}\nn
C_R&=&\int_{-1}^1dx\;\frac{\left(x+v_R\right)^2}{x-\kappa_R}\nn
D_R&=&\int_{-1}^1dx\;x\frac{\left(x+v_R\right)^2}{x-\kappa_R}
\label{ABCDR}
\tea
to the effect that instead of eq. (\ref{param2}) we now get
\be
0=v_R^2-\frac13+v_RG_R\left[\kappa_R\right]
\label{param2R}
\te
where
\be
G_R\left[\kappa_R\right]=\kappa_R-\frac13\frac{J_0\left[\kappa_R\right]}{\kappa_R J_0\left[\kappa_R\right]-2}
\te
$J_0$ as in eq. (\ref{nonanal}). The final parametric relationship between $v_R$ and $\lambda^R_{AW}$ is

\bea
v_R&=&\frac12\left[\sqrt{G_R^2\left[\kappa_R\right]+\frac43}-G_R\left[\kappa_R\right]\right]\nn
\lambda^R_{AW}&=&\frac{T_R}{\tau_0\left(v_R+\kappa_R\right)}\sqrt{1-v_R^2}
\label{param3R}
\tea
Let's analyze the limiting cases. When $\kappa\to\infty$, $G_R\left(\kappa\right)\approx 4/\left(15\kappa\right)$, so $v_R^2\to 1/3$ and $\lambda^R_{AW}\to 0$. 
When $\kappa\to 1$, $J_0\to\infty$, $v_R\to 1/3$,

\be
\lambda^R_{AW}\approx\frac{T_R}{\tau_0}\frac{\sqrt{2}}2
\label{asymp2R}
\te
Finally, let us consider the DTT. The analysis in the main text goes unchanged until eq. (\ref{stop3}), which, after linearization $v=v_R+\vartheta e^{-\lambda^R_{DTT}z}$, becomes (cfr. eq. (\ref{z2v}))

\be
\zeta T_R=\frac56\frac{\left(1-3v_R^2\right)}{v_R\left(1-v_R^2\right)}\vartheta e^{-\lambda^R_{DTT} z}
\label{z2vR}
\te
Also the calculation of the $A$, $B$ and $I$ scalars in eq. (\ref{stop4}) goes unchanged, except that now we linearize around an equilibrium with themperature $T_R$ and velocity $v_R$. Considering that now $\zeta T$ and $\xi T\propto \exp\left\{-\lambda_{DTT}^Rz\right\}$, and writing (cfr. eq. (\ref{lambdadtt}))

\be
\lambda^R_{DTT}=\frac{\alpha^R T_R}{v_R\tau_0}\sqrt{1-v_R^2}
\label{lambdadttR}
\te
we get the set of equations

\be
\left(\begin{array}{ccc}
\frac23\alpha^R &  v_R\left(\alpha^R-1\right)&\frac37\alpha^R\\
-\left(1-3v_R^2\right)& \frac65v_R&0\\
0&\frac35\alpha^R&v_R\left(\alpha^R-1\right)\end{array}\right)\left(\begin{array}{c}\frac{\vartheta}{{1-v_R^2}}\\ \zeta T\\ \xi T\end{array}\right)=0
\te
The allowed values of $\alpha$ are the roots of

\be
a\alpha^{R2}-2bv_R^2\alpha^R+cv_R^2=0
\te
where

\bea
a&=&\frac45v_R^2+\left(1-3v_R^2\right)\left(v_R^2-\frac 9{35}\right)\nn
b&=&\frac7{5}-3v_R^2\nn
c&=&1-3v_R^2
\tea
namely

\be
\alpha=\frac{v_R^2}{a}\left[b-\sqrt{b^2-\frac{ac}{v_R^{2}}}\right]
\label{alphaR}
\te


\begin{thebibliography}{99}

\bibitem{RZ13} L. Rezzolla and O. Zanotti, 
\textit{Relativistic Hydrodynamics} 
(Oxford University Press, Oxford, 2013).


\bibitem{Israel60} W. Israel, Relativistic Theory of Shock Waves, Proc. R. Soc. Lond. A  259, 129 (1960).


\bibitem{Taub73} A. Taub, General Relativistic Shock Waves
in Fluids for which Pressure Equals Energy
Density, Commun. math. Phys. 29, 79 (1973).

\bibitem{Thompson86} K. Thompson, The special relativistic shock tube, J. Fluid Mech. 171,  365 (1986).





\bibitem{CM88} C. Cercignani and A. Majorana, Structure of shock waves in relativistic simple gases, The Physics of Fluids 31, 1064 (1988).

\bibitem{MM90} A. Majorana and O. Muscato, Shock structure in an ultrarelativistic gas, Meccanica 25, 77 (1990).


\bibitem{MM93} J. Mart{\'\i} and E. M\"uller, The analytical solution of the Riemann problem in
relativistic hydrodynamics, J. Fluid Mech.  258, 317 (1994).

\bibitem{KO93} A. Khonkin and A. Orlov, Weak shock structure on the basis of modified hydrodynamical equations, Physics of Fluids A: Fluid Dynamics 5, 1810 (1993).


\bibitem{Bouras09} I. Bouras, E. Moln\'ar, H. Niemi, Z. Xu, A. El, O. Fochler, C. Greiner, and D. H. Rischke, Relativistic Shock Waves in Viscous Gluon Matter, Phys. Rev. Lett. 103, 032301 (2009).

\bibitem{Bouras10} I. Bouras, E. Moln\'ar, H. Niemi, Z. Xu, A. El, O. Fochler, C. Greiner, and D. H. Rischke, Investigation of shock waves in the relativistic Riemann problem: A comparison of viscous fluid
dynamics to kinetic theory, Phys. Rev. C 82, 024910 (2010).

\bibitem{MNR10} E. Moln\'ar, H. Niemi and D.H. Rischke, Numerical tests of causal relativistic dissipative fluid dynamics, Eur. Phys. J. C65, 615 (2010).

\bibitem{KKM10} S. Khlebnikov, M. Kruczenski and G. Michalogiorgakis, Shock waves in strongly coupled plasmas, Phys. Rev. D 82, 125003 (2010).


\bibitem{Herbst19} R. Herbst, A Review of the Relativistic Euler Gas Equations: The
Numerical Intricacies, AIP Conference Proceedings 2116, 030037 (2019).

\bibitem{RXZ20} T. Ruggeri, Q. Xiao, H. Zhao, The Riemann problem of relativistic Euler system with Synge energy, arXiv:2001.04128v1.

\bibitem{LO11} X. Liu and Y. Oz, Shocks and universal statistics in (1+1)-dimensional
relativistic turbulence, JHEP 03, 006 (2011).

\bibitem{ED18} G. Eyink and Th. Drivas, Cascades and Dissipative Anomalies in Relativistic Fluid Turbulence, Phys. Rev. X 8, 011023 (2018).

\bibitem{HF21} H. Freist\"uhler, Nonexistence and existence of shock profiles in the
Bemfica-Disconzi-Noronha model, Phys. Rev. D 103, 124045 (2021).

\bibitem{Gabbana20} A. Gabbana, S. Plumari, G. Galesi, V. Greco, D. Simeoni, S. Succi, and R. Tripiccione, Dissipative hydrodynamics of relativistic shock waves in a Quark Gluon Plasma: comparing and benchmarking alternate numerical methods, Phys. Rev. C 101 064904 (2020).

\bibitem{I88} W. Israel, 
Covariant fluid mechanics and thermodynamics: An introduction,
in \textit{Relativistic Fluid Dynamics}, edited by A. M. Anile and Y. Choquet-Bruhat
(Springer, New York, 1988), p. 152.




\bibitem{OH90} T. Olson and W. Hiscock, Plane Steady Shock Waves in Israel-Stewart Fluids, Ann. Phys. 204, 331 (1990).


\bibitem{Muller99} I. M\"uller, Speeds of propagation in classical and relativistic
extended thermodynamics, Living Rev. Relativity 2, 1
(1999).

\bibitem{WCU70} C. S. Wang Chang and G. E. Uhlenbeck, The kinetic theory of gases, in \emph{Studies in Statistical Mechanics}, Vol. 5, edited by J. De Boer and G. E. Uhlenbeck (North Holland, Amsterdam, 1970), p. 1.

\bibitem{Sir63a} L. Sirovich, Dispersion Relations in Rarefied Gas Dynamics, Physics of Fluids (1958-1988) 6, 10 (1963).

\bibitem{Sir63b} L. Sirovich, Formal and Asymptotic Solutions in Kinetic Theory, Physics of Fluids (1958-1988) 6, 218 (1963).

\bibitem{Strut08} H. Struchtrup, Linear Kinetic Heat Transfer: Moment Equations, Boundary Conditions, and Knudsen layers, Physica A 387, 1750 (2008).

\bibitem{J14} B. M. Johnson, Closed-form shock solutions, J. Fluid Mech. 745, R1 (2014).


\bibitem{StrutBook} H. Struchtrup, \emph{Macroscopic Transport Equations for Rarefied Gas Flows} (Springer, Berlin, 2005).


\bibitem{MJ89} V. Micenmacher and D. Jou, On the convexity of a non equilibrium entropy and shock waves, Phys. Lett. A 141, 165 (1989).

\bibitem{JP91} D. Jou and D. Pav\'on, Nonlocal and nonlinear effects in shock waves, Phys. Rev. A 44, 6496 (1991).

\bibitem{DMNR11} B. Betz, G.S. Denicol, T. Koide, E. Moln\'ar, H. Niemi, and D.H. Rischke, Second order dissipative fluid dynamics from kinetic theory, Eur. Phys. J.  Conf.13:07005,2011

\bibitem{DMNR11b} G.S. Denicol, J. Noronha, H. Niemi, and D.H. Rischke,   Origin of the relaxation time in dissipative fluid dynamics, Phys. Rev. D 83, 074019 (2011)

\bibitem{DMNR12}  G. S. Denicol, E. Moln\'ar, H. Niemi and D. H. Rischke, 
Derivation of fluid dynamics from kinetic theory with the 14 moment approximation, 
Eur. Phys. J. A 48 11 (2012).

\bibitem{DMNR12b} G. S. Denicol, H. Niemi, E. Moln\'ar and D. H. Rischke
Derivation of transient relativistic fluid dynamics from the Boltzmann equation,
Phys. Rev. D 85, 114047 (2012); Phys. Rev. D 91, 039902(E) (2015).

\bibitem{BR99} G. Boillat and T. Ruggeri, Relativistic gas: Moment equations
and maximum wave velocity, J. Math. Phys. (N.Y.) 40,
6399 (1999).

\bibitem{RR19} P. Romatschke and U. Romatschke, 
\textit{Relativistic fluid dynamics in and out equilibrium - Ten years of progress in theory and numerical simulations of nuclear collisions} (Cambridge University Press, Cambridge (England), 2019).


\bibitem{AW1} J. L. Anderson and H. R. Witting, A Relativistic Relaxation-Time Model for the Boltzmann Equation, Physica 74, 466 (1974).
	
\bibitem{AW2} J. L. Anderson and H. R. Witting,  Relativistic Quantum Transport Coefficients, Physica 74, 489 (1974).

\bibitem{TI10} M. Takamoto and S. I. Inutsuka, The relativistic kinetic dispersion relation: 
Comparison of the relativistic Bhatnagar$-$ Gross$-$ Krook model and Grad's 14-moment expansion,
Physica A, 389, 4580 (2010).

\bibitem{KW19} A. Kurkela and U. A. Wiedemann, Analytic structure of
nonhydrodynamic modes in kinetic theory, Eur. Phys. J. C
79, 776 (2019).

\bibitem{PC21} G. Perna and E. Calzetta, Linearized dispersion relations in viscous relativistic hydrodynamics, Phys. Rev. D 104, 096005 (2021)

\bibitem{L72} I. S. Liu, Method of Lagrange Multipliers for Exploitation of the Entropy Principle, Arch. for Rat. Mech. and Anal. 46, 2, 131 (1972).

\bibitem{LMR86} I-S. Liu, I. M\"uller and T. Ruggeri, Relativistic Thermodynamics of Gases, Annals of Physics 169, 191 (1986).

\bibitem{GL90} R. Geroch and L. Lindblom, 
Dissipative relativistic fluid theories of divergence type,
Phys. Rev. D 41,  1855 (1990). 

\bibitem{GL91} R. Geroch and L. Lindblom, 
Causal theories of dissipative relativistic fluids,
Ann. Phys. (NY) {207}, 394 (1991).

\bibitem{PRC09} J. Peralta-Ramos and E. Calzetta, Divergence-type nonlinear conformal hydrodynamics, Phys. Rev. D 80, 126002 
(2009).

\bibitem{PRC10a} E. Calzetta and J. Peralta-Ramos,
Linking the hydrodynamic and kinetic description of a dissipative relativistic conformal theory,
Phys. Rev. D82, 106003 (2010)

\bibitem{PRC10b} J. Peralta-Ramos and E. Calzetta,
Divergence-type 2+1 dissipative hydrodynamics applied to heavy-ion collisions,
Phys. Rev. C82, 054905 (2010)

\bibitem{PRC13a} J. Peralta-Ramos and E. Calzetta,
Macroscopic approximation to relativistic kinetic theory from a nonlinear closure,
Phys. Rev. D 87, 034003 (2013).

\bibitem{lucas19} L. Cantarutti and E. Calzetta, 
Dissipative-type theories for Bjorken and Gubser flows, 
Int. J. Mod. Phys. A 35, 2050074 (2020).

\bibitem{MGC17} N. Mir\'on Granese and E. Calzetta, 
Primordial gravitational waves amplification from causal fluids,
Phys. Rev. D {97}, 023517  (2018).

\bibitem{MG21}  N. Mir\'on Granese,
Relativistic viscous effects on the primordial gravitational waves spectrum,
JCAP 06, 008 (2021).

\bibitem{MGKC20} N. Mir\'on-Granese, A. Kandus, and E. Calzetta, Nonlinear
fluctuations in relativistic causal fluids, JHEP 07, 064 (2020).

\bibitem{EC21} E. Calzetta, Fully developed relativistic turbulence, Phys.
Rev. D 103, 056018 (2021).

\bibitem{MGKC21} N. Mir\'on-Granese, A. Kandus, and E. Calzetta, Primordial Weibel instability,  JCAP 01, 028 (2022).

\bibitem{LL6} L. D. Landau and E. M. Lifshitz, \textit{Fluid Mechanics}  (Pergamon Press, Oxford, England, 1959).

\bibitem{IS79a} W. Israel and J. M. Stewart, Transient Relativistic Thermodynamics and Kinetic Theory, Annals of Physics 118, 341 (1979).

\bibitem{IS79b} W. Israel and M. Stewart, 
On transient relativistic thermodynamics and kinetic theory. II,
Proc. R. Soc. London, Ser A 365, 43 (1979).

\bibitem{MM85} A. Majorana and S. Motta, Shock Structure in Relativistic Fluid-Dynamics, J. Non-Equilib. Thermodyn.
10, 29 (1985).

\bibitem{DKKM08} G. S. Denicol, T. Kodama, T. Koide, and Ph. Mota, Shock propagation and stability in causal dissipative hydrodynamics, Phys. Rev. C 78, 034901 (2008).

\bibitem{U16} F. J. Uribe, Shock waves: The Maxwell-Cattaneo case, Phys. Rev. E 93, 033110 (2016).


\bibitem{ST14} M. Strickland, 
Anisotropic Hydrodynamics: Three lectures, 
Act. Phys. Pol. B 45, 2355 (2014).

\bibitem{KKL19} M. Kirakosyan, A. Kovalenko, and A. Leonidov, Sound propagation and Mach cone in anisotropic
hydrodynamics, Eur. Phys. J. C 79, 434 (2019).


\bibitem{KL21} A. Kovalenko and A. Leonidov, Shock Waves in Relativistic Anisotropic Hydrodynamics, arXiv:2103.06745v2 [
nucl-th] (2021).




\bibitem{L08} R. Loganayagam, 
Entropy current in conformal hydrodynamics,
JHEP 05, 087 (2008).

\bibitem{JBP13} A. Jaiswal, R. Bhalerao and S. Pal, Complete relativistic second-order dissipative hydrodynamics from the entropy principle, Phys. Rev. C 87, 021901 (2013).

\bibitem{CJPR15} Ch. Chattopadhyay, A. Jaiswal, S. Pal, and R. Ryblewski, Relativistic third-order viscous corrections to the entropy four-current from kinetic theory, Phys. Rev. C 91, 024917 (2015).

\bibitem{DMNR10} G.S. Denicol, T. Koide, and D.H. Rischke, Dissipative relativistic fluid dynamics: a new way to derive the equations of motion from kinetic theory, Phys. Rev. Lett. 105, 162501 (2010).




\end{thebibliography}
\end{document}